\documentclass[11pt,preprint]{aastex}
\newcommand{\iras}{IRAS 04158+2805 }

\begin{document}

\shorttitle{Submillimeter Observations of IRAS 04158+2805}

\shortauthors{Andrews et al.}

\title{Submillimeter Observations of the Young Low-Mass Object \\ 
IRAS 04158+2805}

\author{Sean M. Andrews\altaffilmark{1,2}, 
Michael C. Liu\altaffilmark{3,4}, Jonathan P. Williams\altaffilmark{3}, and K. 
N. Allers\altaffilmark{3}}

\altaffiltext{1}{Harvard-Smithsonian Center for Astrophysics, 60 Garden 
Street, Cambridge, MA 02138; sandrews@cfa.harvard.edu}
\altaffiltext{2}{Hubble Fellow}
\altaffiltext{3}{Institute for Astronomy, University of Hawaii, 2680 Woodlawn 
Drive, Honolulu, HI 96822; mliu@ifa.hawaii.edu, jpw@ifa.hawaii.edu, allers@ifa.hawaii.edu}
\altaffiltext{4}{Alfred P. Sloan Research Fellow}

\begin{abstract}
We present high spatial resolution Submillimeter Array observations and 
supplementary single-dish photometry of the molecular gas and dust around IRAS 
04158+2805, a young source with spectral type M5-M6 in the Taurus star-forming 
region.  A bright, highly elongated dust structure that extends 8\arcsec\ 
($\sim$1120\,AU) in diameter is revealed in a 883\,$\mu$m thermal continuum 
image.  This emission geometry is in good agreement with optical observations 
that show a similar structure in absorption, aligned perpendicular to bipolar 
scattered light nebulae.  However, the interferometric data also clearly 
demonstrate that the submillimeter continuum emission is not centrally 
concentrated, but rather appears to have a toroidal geometry with substantially 
lower intensities inside a radius of $\sim$250-300\,AU.  Spatially resolved 
emission from the CO $J$=3$-$2 transition exhibits a velocity gradient along 
the major axis of the dust structure.  If this kinematic pattern is interpreted 
as the signature of rotation around a central object, a relatively low mass is 
inferred ($M_{\ast} \sim 0.3$\,M$_{\odot}$, with a $\sim$50\% uncertainty).  We 
discuss several possible explanations for the observed gas and dust environment 
around IRAS 04158+2805, including a flattened envelope with an outflow cavity 
and a large circumbinary ring.  This source offers unique views of the gas and 
dust environment surrounding a young low-mass stellar system.  Its properties 
are generally not commensurate with formation scenarios for such low-mass 
objects that rely on dynamical ejection, but rather confirms that a single 
mechanism $-$ molecular cloud core collapse and fragmentation $-$ can produce 
stars over a wide range of stellar masses (at least an order of magnitude).
\end{abstract}


\section{Introduction}

Circumstellar material profoundly influences the star formation process.  A 
large-scale envelope acts as the local mass reservoir during the collapse and 
growth of a central protostar; a more compact disk regulates how that material 
is transported onto the star itself.  A great deal of progress has been made in 
understanding the physical conditions present in the gas and dust surrounding 
Sun-like stars ($M_{\ast} \approx 0.5$-3\,M$_{\odot}$).  However, it is unclear 
if those results can be extrapolated across the entire stellar mass spectrum.  
The low end of this spectrum ($M_{\ast} \le 0.3$\,M$_{\odot}$) is of particular 
interest, as observations of the material around young low-mass 
objects\footnote{For simplicity, we will refer to objects with $M_{\ast} 
\lesssim 0.3$\,M$_{\odot}$ as low-mass objects, including both very low-mass 
stars and brown dwarfs.} can help resolve an ongoing debate over their dominant 
formation mechanism.

The observed abundance of such low-mass objects relative to their more massive 
counterparts is difficult to account for with the traditional model for 
isolated star formation.  Two distinct modifications have been proposed: ($a$) 
the turbulent fragmentation of cloud cores into smaller building units 
\citep[e.g.,][]{padoan04}; and ($b$) a dynamical alternative where one 
component of a multiple system is ejected and prematurely cut off from its 
accretion reservoir \citep[e.g.,][]{reipurth01,umbreit05}.  Both scenarios make 
distinct and conflicting predictions about the gas and dust environments that 
should be associated with low-mass objects.  In the former, a disk/envelope 
structure similar to those noted around higher mass T Tauri stars would be 
expected.  And for the latter, only a truncated disk (with an initial $R_d 
\lesssim 10$\,AU) would survive the tidal stripping and ejection process 
\citep{bate03}.  If the remnant material is sufficiently viscous, it could 
spread to radii up to $\sim$100\,AU in $\sim$1\,Myr \citep{armitage97}, but 
would have shed a vast majority of its initial mass in the ejection process.  
Therefore, observational constraints on the structure of the gas and dust 
around low-mass objects offer one avenue to help distinguish how they form.

A variety of observations demonstrate that the signatures of circumstellar gas 
and dust at small radii (up to a few AU) are indeed common for young low-mass 
objects \citep[see the recent review by][]{luhman07b}.  More detailed 
individual studies confirm that such material has a geometry and composition 
similar to the disks around more massive T Tauri stars 
\citep{natta01,pascucci03,allers06,buoy08}, although typically at lower masses 
\citep{klein03,scholz06}.  However, these {\it unresolved} observations can not 
unambiguously constrain some key properties of this material, most 
significantly its spatial structure and extent.

In this article, we present new submillimeter observations of the molecular gas 
and dust surrounding IRAS 04158+2805.  Located in the $\sim$1\,Myr-old Taurus 
star-forming region, \iras has a cool central source with an estimated spectral 
type of M5-M6 \citep{white04,luhman06,beck07}.  Recently, \citet{glauser08} 
provided an initial analysis of this source and the dust structure surrounding 
it by modeling the broadband spectrum along with optical/infrared scattered 
light images.  They suggested that their observations are best explained by a 
large (diameter of 2240\,AU), massive (dust mass of 
1-2$\times$10$^{-4}$\,M$_{\odot}$; gas/dust ratio of $220^{+150}_{-170}$) 
circumstellar disk around a low-mass star ($M_{\ast} \approx 
0.1$-0.2\,M$_{\odot}$), with no need to include any contribution from a 
more extended envelope.  Our new data represent the first high spatial 
resolution view of the gas and dust environment around this and similar cool, 
young objects at submillimeter wavelengths.  They also provide an opportunity 
for a rare, albeit crude, estimate of $M_{\ast}$ from the spatio-kinematics of 
circumstellar gas that is independent of pre-main-sequence evolution models.  
These observations are addressed in \S 2, and the resulting data products are 
highlighted in \S 3.  In \S 4, we discuss the structure of the gas and dust 
surrounding \iras and what it can tell us about the central source and the 
formation of low-mass objects in general.  The results are summarized in \S 5.

\section{Observations and Data Reduction}

Submillimeter interferometric observations of \iras ($\alpha = 
4^{\mathrm h}18^{\mathrm m}58\fs14$, $\delta = +28\degr12\arcmin23\farcs8$ 
[J2000]) were conducted in both the extended and compact configurations of the 
Submillimeter Array \citep[SMA;][]{ho04} on 2006 December 8 and 2007 January 
27, respectively.  In these configurations, the 8 SMA antennas (6\,m diameter 
each) spanned baselines from $\sim$15-225\,m.  Double sideband receivers with a 
total bandwidth of 4\,GHz were tuned to an effective continuum frequency of 
339.9\,GHz (883\,$\mu$m).  The correlator was simultaneously configured to 
observe the CO $J$=3$-$2 transition at 345.796\,GHz at a spectral resolution of 
0.70\,km s$^{-1}$.  

The observations cycled between \iras and two gain calibrators (3C111 and 
3C84), with 20 minutes on target and 10 minutes on calibrator.  The data were 
obtained in excellent observing conditions, with zenith optical depths at 
225\,GHz of 0.05-0.07.  The raw visibilities were calibrated with the IDL-based 
MIR package.\footnote{\url{http://cfa-www.harvard.edu/$\sim$cqi/mircook.html.}} 
Passband calibration was conducted with bright quasars (3C273, 3C279), and 
complex gain calibration was performed with 3C111.  Using 3C84 as a consistency 
check on the gain solution, we estimate that phase noise generates an effective 
``seeing" of $<$0\farcs1.  Titan and Callisto were used to set the absolute 
flux scale, which is accurate to $\sim$10\%.  The MIRIAD package was utilized 
for the standard tasks of Fourier inversion, deconvolution, and imaging of the 
calibrated visibilities.  Continuum and line maps from the combined datasets 
were made with natural weighting, providing a synthesized beam FWHM of 
$1\farcs12\times0\farcs86$ at a position angle of 92\degr.

\iras was also mapped at 450 and 850\,$\mu$m using the jiggle-mode of the SCUBA 
camera at the 15\,m James Clerk Maxwell Telescope (JCMT) on 2003 November 12 in 
dry, stable conditions ($\tau \approx 0.05$ at 225\,GHz).  Flux calibration 
accurate to $\sim$10\%\ and 25\%\ at 850 and 450\,$\mu$m, respectively, was 
achieved with observations of Neptune, and pointing was referenced to DG Tau 
and CRL 618.  The bright emission, $0.44\pm0.04$\,Jy at 850\,$\mu$m and 
$1.6\pm0.4$\,Jy at 450\,$\mu$m, was not resolved at either wavelength in 
15\arcsec\ and 9\arcsec\ beams, respectively.  An additional map at 
350\,$\mu$m was obtained with the SHARC-II camera at the 10\,m Caltech 
Submillimeter Observatory (CSO) on 2004 October 3 in similarly dry weather.  
Those data confirm an unresolved source (9\arcsec\ beam) with a total flux 
density of $2.0\pm0.5$\,Jy.  All errors on flux densities include absolute 
calibration uncertainties (10\%\ at 850\,$\mu$m, and $\sim$25\%\ at shorter 
wavelengths).  

Optical images of \iras in the $R$-band were obtained with the University of 
Hawaii 2.2\,m telescope on 2004 October 6.  After the standard reduction of CCD 
data, the $7\times300$\,s images were co-added and convolved with a 3-pixel 
(0\farcs7) FWHM Gaussian kernel to match the typical seeing for the 
observations and improve sensitivity to faint emission.  The observations were 
not photometric.

\section{Results}

\begin{figure}
\epsscale{1.}
\plotone{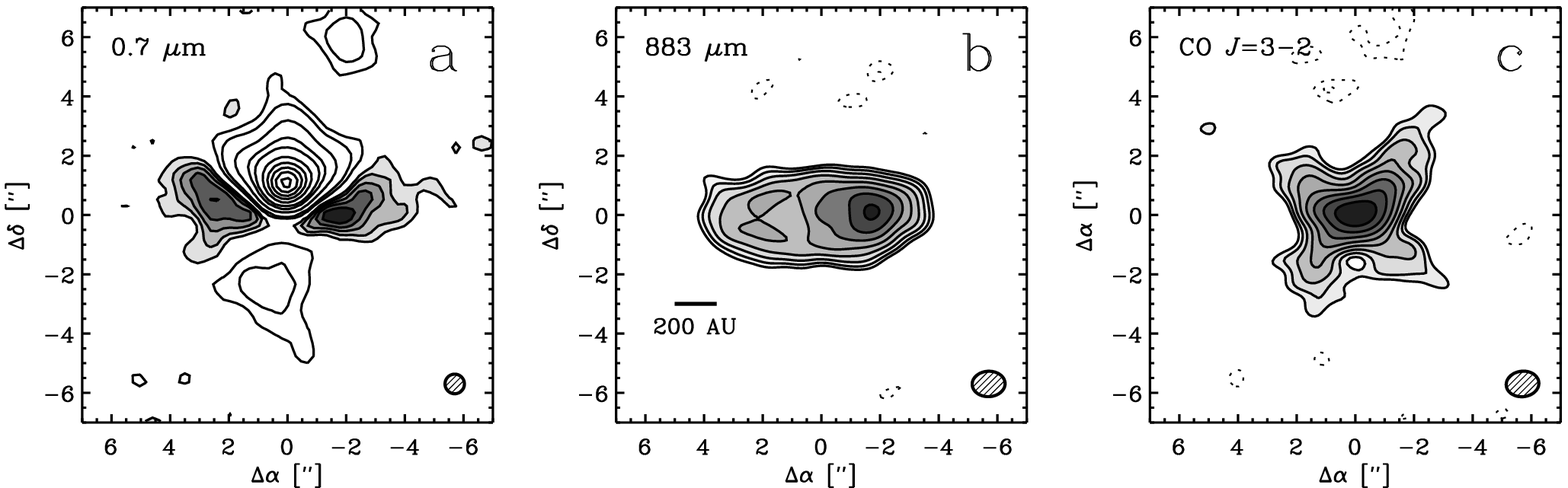}
\caption{Images of the gas and dust surrounding IRAS 04158+2805.  ({\it left}) 
Optical $R$-band (0.7\,$\mu$m) image showing a dark lane in silhouette 
(grayscale contours; linear steps) against background nebulosity and scattered 
starlight (unfilled contours; logarithmic steps).  ({\it middle}) Submillimeter 
(883\,$\mu$m) dust continuum image from the SMA.  Contours start at 3\,$\sigma$ 
($\sim$3.5\,mJy beam$^{-1}$) and increase by factors of $\sqrt{2}$.  ({\it 
right}) Velocity-integrated (0$^{\mathrm{th}}$ moment) emission image from the
CO $J$=3$-$2 transition, constructed from 9 spectral channels (a velocity 
width of 6.3\,km s$^{-1}$).  Contours start at 3\,$\sigma$ (0.5\,Jy km s$^{-1}$ 
beam$^{-1}$) and increase by factors of $\sqrt{2}$.  Effective FWHM 
(synthesized) beam sizes are shown in the lower right corner of each panel.
\label{images}}
\end{figure}

Images of the gas and dust surrounding \iras are shown in Figure \ref{images}.  
The optical $R$-band ($\sim$0.7\,$\mu$m) image in Figure \ref{images}a reveals 
a high aspect ratio dark lane of {\it absorbing} material oriented roughly E-W 
(shown as grayscale-filled contours) that can be traced nearly 8\arcsec\ in 
diameter.  First discovered by \citet[][see also Watson et 
al.~2007]{glauser08}, this silhouette lane obscures the faint nebulosity 
associated with the nearby Herbig Ae star V892 Tau.  Also pointed out by 
\citet{glauser08}, the image in Figure \ref{images}a shows the conical nebulae 
perpendicular to the dark lane that are often produced by starlight scattered 
off the surface of flattened dust structures. 

Figure \ref{images}b shows the first high resolution image of the thermal 
continuum emission from IRAS 04158+2805, at a wavelength of 883\,$\mu$m.  The 
morphology of that emission is remarkably similar to the optical silhouette, 
extending roughly 8\arcsec\ across, corresponding to a diameter of 1120\,AU at 
the distance of the Taurus clouds \citep[$d \approx 140$\,pc;][]{elias78}.  
Note that \citet{glauser08} trace $I$-band scattered light for this source to 
twice this diameter ($\sim$16\arcsec, or 2240\,AU).  The difference compared to 
the $R$-band image in Figure \ref{images}a is related to the comparatively poor 
sensitivity of our observations, as was also noted for the infrared scattered 
light images in their study.  Because the submillimeter emission was clearly 
unresolved in the 350 and 450\,$\mu$m maps, we can infer that it has a diameter 
of $\le 9\arcsec$ (1260\,AU).  The absence of very short antenna spacings 
implies that the interferometric observations have significantly diminished 
sensitivity to structures larger than $\sim$10\arcsec\
\citep[e.g.,][]{wilner94}.  Therefore, because of the diminished sensitivity to 
faint submillimeter emission on large scales, there is no reason to suspect 
that the sized inferred here is physically inconsistent with the scattered 
light image presented by \citet{glauser08}.  The integrated flux density at 
883\,$\mu$m is $407\pm41$\,mJy, consistent with the SCUBA measurement.  A fit 
of the SMA visibilities with an elliptical Gaussian model indicates an 
inclination of $62\pm3\degr$ (where 90\degr\ corresponds to edge-on) at a 
position angle of $93\pm1$\degr\ (measured east of north), in excellent 
agreement with the constraints imposed by modeling the scattered light 
\citep{glauser08}.  Fits with a ring or pedastal emission model give the same 
results.  

\begin{figure}[b]
\epsscale{0.6}
\plotone{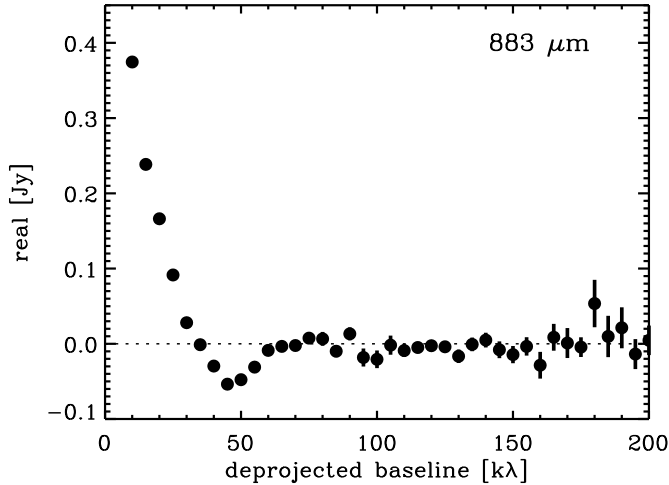}
\caption{The real part of the \iras 883\,$\mu$m visibilities as a function of
the interferometer baseline, after deprojection and elliptical averaging
according to the observed geometry \citep[see][]{lay97}.  The steep drop in the 
visibilities at short baselines and the pronounced null at 
$\sim$30-40\,k$\lambda$ confirm the morphology in Fig.~1b; the submillimeter 
emission is not centrally peaked.  \label{vis}}
\end{figure}

The high surface brightness at large distances from the central source and 
large spatial extent of the continuum emission are fairly unique.  The steep 
drop of the visibilities over a short range of baseline lengths, as shown in 
Figure \ref{vis}, indicates that essentially all of the emission is 
concentrated at large spatial scales (radii $\gtrsim 250$\,AU).  More 
significantly, the visibilities show a pronounced null at 
$\sim$30-40\,k$\lambda$ and very little flux on longer baselines.  These are 
definitive signatures of a brightness distribution with a central ``hole", 
inside of which there is significantly diminished intensity.  Assuming a simple 
ring geometry, the location of the null suggests an inner radius of 
$\sim$250-300\,AU \citep[e.g.,][]{hughes07}.  Asymmetries in the submillimeter 
map confirm this geometry, with emission peaks centered roughly $\pm 1\farcs9$ 
($\sim$265\,AU) from the image center.  The western lobe is approximately 50\%\ 
brighter than the eastern lobe.  These asymmetries are real, as they are 
clearly detected in 3 separate observing runs at the same location and 
intensity \citep[the 2 detailed here, and the hybrid compact/extended 
observations described by][]{aw07}\footnote{In fact, the image made from the 
hybrid configuration in \citet{aw07} shows {\it only} the western peak of this 
asymmetry, and heavily resolves the remainder of the emission.}.

\begin{figure}
\epsscale{1.05}
\plotone{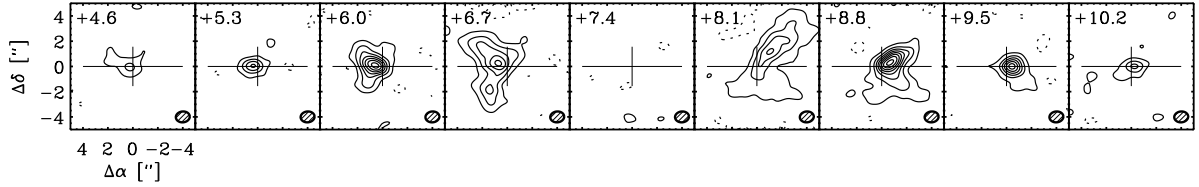}
\caption{Channel maps of the \iras CO $J$=3$-$2 line emission.  Contours start
at 3\,$\sigma$ (0.2\,Jy beam$^{-1}$) and increase linearly in 6\,$\sigma$ 
increments.  The FWHM synthesized beam sizes are shown in the lower right of 
each panel, and the LSR velocity of each channel (in km s$^{-1}$) is shown in 
the upper left.  The crosshairs mark the horizontal extent of the submillimeter 
continuum emission and the vertical extent corresponding to the inclination 
(62\degr) derived from fits of the continuum visibilities.  There is a clear 
velocity gradient along the major axis of the continuum emission and optical 
silhouette.  
\label{chanmaps}}
\end{figure}

A velocity-integrated intensity (0$^{\mathrm{th}}$ moment) map of the CO 
$J$=3$-$2 emission from \iras is shown in Figure \ref{images}c.  The CO 
emission morphology is significantly different than the optical silhouette and 
submillimeter continuum, occupying a region $\sim$4\arcsec\ on a side with a 
central, slightly elongated peak.  The integrated intensity is $40.7\pm0.4$\,Jy 
km s$^{-1}$ (using 9 channels for a total width of 6.3\,km s$^{-1}$).  Fainter 
CO emission extends above and below the plane of the continuum emission in an 
``X"-shaped pattern out to $\sim$2\arcsec\ from the image center.  There is a 
steep drop in the CO intensity along the plane of the silhouette that 
corresponds precisely to the positions of the submillimeter continuum peaks 
described above.  The kinematic structure of the CO associated with \iras is 
exhibited in Figure \ref{chanmaps}.  These channel maps clearly show a CO 
velocity gradient along the major axis of the continuum emission and silhouette 
(i.e., in the east-west direction), similar to the pattern expected for 
rotation around a central source.

\section{Discussion}

\subsection{The Central Object}

Estimating the masses of young stars and brown dwarfs generally requires 
reference to theoretical models of their structural evolution.  The 
quantitative reliability of such models for individual stars at ages 
$\lesssim$1\,Myr is highly uncertain, particularly for the low end of the 
stellar mass spectrum where complicated convection physics, dusty atmospheres, 
and uncertain molecular opacities and initial conditions can strongly affect 
the observational diagnostics 
\citep[e.g.,][]{dantona94,baraffe02,montalban04}.  Dynamical constraints on 
$M_{\ast}$ from the orbital properties of either a companion star 
\citep[e.g.,][]{mathieu94} or gas in a circumstellar disk 
\citep[e.g.,][]{simon00} provide an extremely valuable independent check that 
can potentially be used to calibrate these models 
\citep[see][]{hillenbrand04}.  \citet{stassun06} report the only such 
measurements of young $\sim$M6 dwarfs to date, for an eclipsing binary system 
in Orion.  With the first spatially resolved measurements of molecular gas in 
apparent rotation around a young low-mass object with similar spectral type, 
the CO observations of \iras presented here can provide another such estimate 
of $M_{\ast}$, although admittedly with a great deal more uncertainty.

\begin{figure}
\epsscale{0.5}
\plotone{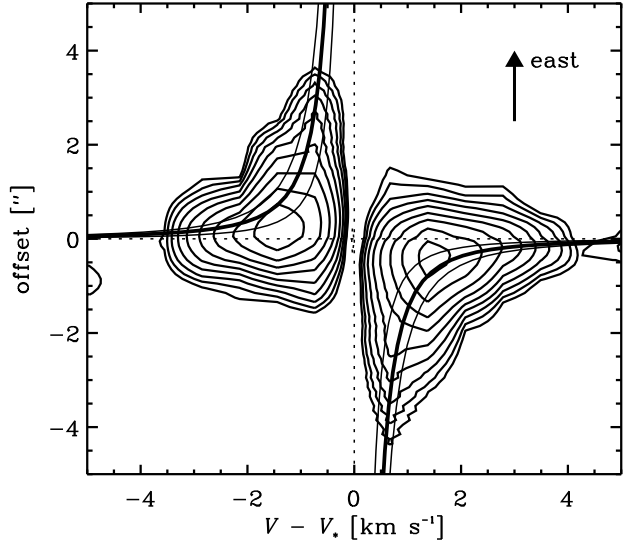}
\caption{Position-velocity diagram of the CO $J$=3$-$2 line emission.  Contours
start at 0.1\,Jy and increase in steps of $\sqrt{2}$.  The ordinate shows the
projected offset from the image center (with positive values to the east), and 
the abscissa shows the velocity offset from the systemic value, +7.4\,km 
s$^{-1}$.  The heavy curve represents an inclined ($i = 62$\degr) Keplerian 
rotation profile that is most similar to the data, for $M_{\ast} = 
0.3$\,M$_{\odot}$.  Lighter curves on either side show similar profiles for 
$M_{\ast} = 0.15$ and 0.45\,M$_{\odot}$.  \label{pv_spec}}
\end{figure}

Figure \ref{pv_spec} shows the position-velocity diagram of the CO $J$=3$-$2 
line emission for IRAS 04158+2805, where the angular offset is the distance 
from the observed phase center (such that positions to the east have positive 
values) and the velocity offset is taken relative to the systemic value 
($V_{\ast} = 7.4$\,km s$^{-1}$).  Overlaid on the diagram is an inclined ($i = 
62$\degr) Keplerian rotation profile that matches the general kinematic trend 
in the data rather well (thick curve), with a central point mass of $M_{\ast} = 
0.30$\,M$_{\odot}$.  Additional rotation profiles for $M_{\ast} = 0.15$ and 
0.45\,M$_{\odot}$ are also shown for reference.  It should be noted that these 
profiles assume pure Keplerian orbital motion around a central source, which 
may not be the case in the more complex \iras environment.  For now, we assume 
that the dominant kinematic trend is such rotation, but potential complications 
will be addressed below.

A comparison of this dynamical $M_{\ast}$ estimate with predictions from 
stellar evolution models is a challenge because of uncertainties in the 
luminosity (due to scattered light contamination at short wavelengths), 
extinction, and effective temperature for the central source.  Recent analysis 
of the scattered light spectrum from \iras suggest that this source is cool, 
with initial spectral type estimates of M6$\pm1$ \citep{white04,beck07} refined 
to M5.25$\pm$0.25 (optical) and M6$\pm$0.5 
\citep[infrared;][]{luhman06}.\footnote{Note that previous spectral type 
assignments ranged from $\sim$K7 to M3 \citep{kenyon98,luhman99}.}  Using the 
latter classifications and the \citet{luhman03} empirical effective temperature 
scale, we estimate $T_{\ast} = 3050\pm75$\,K.  Unfortunately, large 
uncertainties in the extinction, where estimates range from $A_V \approx 9$-16 
\citep[e.g.,][]{white04,beck07}, make a luminosity determination difficult.  
However, we can use pre-main-sequence models for an assumed age in the inferred 
temperature range to estimate a central mass.  Using the $T_{\ast}$ range 
quoted above and an age of $\sim$1\,Myr, the \citet{dantona97} and 
\citet{baraffe98} models indicate $M_{\ast} \approx 0.09$-0.16\,M$_{\odot}$.  
This mass range does not include the additional uncertainties on the adopted 
effective temperature scale or the unknown age of the source.  

The $M_{\ast}$ values inferred from the pre-main-sequence models lie roughly a 
factor of 2 below the nominal value that best describes the kinematics of the 
CO gas inferred above.  The unquoted and poorly understood uncertainties in 
both measurements may be the simple explanation for this apparent discrepancy.  
Without a detailed physical model for the local gas structure around IRAS 
04158+2805, it is difficult to estimate a statistically acceptable range of 
$M_{\ast}$ from the CO kinematics; there may very well be overlap with the 
higher end of the $M_{\ast}$ range from the pre-main-sequence models.  The 
effective temperature scale adopted above is also uncertain.  A $\sim$500\,K 
increase in the effective temperature (to a value that is typically associated 
with M2 stars) would reconcile the $M_{\ast}$ estimates.  Indeed, using an 
independent method of fitting infrared spectra, \citet{doppmann05} infer 
$T_{\ast} = 3500$\,K for the \iras central source, leading to $M_{\ast} \approx 
0.35$-0.45\,M$_{\odot}$ in the aforementioned models.  Moreover, 
\citet{hillenbrand04} have shown that pre-main-sequence models generally tend 
to underpredict $M_{\ast}$ compared to dynamical measurements in this low-mass 
range, although typically only by $\le 20$\%.  In their recent study that 
explicitly includes the effects of scattered starlight, \citet{glauser08} argue 
for a very luminous central source, $L_{\ast} \approx 0.4$\,L$_{\odot}$.  While 
this value must be somewhat degenerate with the dust properties and structure 
assumed in their model, it is clearly the most robust estimate available.  The 
pre-main-sequence models imply that such a luminous cool source would be very 
young, with an age significantly less than 1\,Myr.  The model mass tracks at 
such ages are not well understood in light of the uncertain initial conditions.

Clearly, the aforementioned uncertainties can be solely responsible for the 
apparently different $M_{\ast}$ estimates from the CO kinematics and 
pre-main-sequence models.  Alternatively, there is also an appealing and simple 
physical explanation $-$ the \iras central source may be a roughly equal-mass 
binary.  In this scenario, the $M_{\ast}$ estimate from the CO kinematics 
refers to the total mass of the stellar system (i.e., the mass interior to the 
gas that produces the line emission), while the estimate from the 
pre-main-sequence models assumes only a single star.  Therefore we would expect 
to see both a high luminosity \citep[as claimed by][]{glauser08} and 
$M_{\ast}$(CO) $\sim 2 M_{\ast}$(models).  A variety of surveys suggest that 
such binaries are fairly common, $\sim$35-45\% for M dwarfs 
\citep{fischer92,reid97} and $\sim$10-30\% for cooler objects 
\citep[e.g.,][]{burgasser07}.  While it is difficult to definitively state the 
properties of the central source(s), we can further assess the nature of \iras 
by analyzing the structure of the material that surrounds it.

\subsection{The Gas and Dust Environment}

Given the elongated structures shown in Figure \ref{images}, it is reasonable 
to suggest that the gas and dust around \iras reside in an exceptionally large 
circumstellar disk.  \citet{glauser08} have recently demonstrated that they are 
able to fit high-quality scattered light images and the SED for this source 
with a simple disk structure, and without the need to invoke any envelope 
component.  With the new data presented here, it is worthwhile to re-examine 
the structure of this material.  

\iras is exceptionally bright at submillimeter wavelengths, with an 850\,$\mu$m 
luminosity larger than $\sim$85\% of all other sources detected in the 
\citet{aw05} Taurus survey.  Breaking that sample down further, the \iras 
submillimeter emission is brighter than $\sim$90\% of Class II sources (disk 
only), but is comparable to the median Class I source (disk + envelope).  If we 
adopt the standard emissivity for disks (0.034\,cm$^2$ g$^{-1}$ at 883\,$\mu$m, 
including a gas-to-dust ratio of 100) and assume that the submillimeter 
continuum emission is optically thin and has a characteristic temperature of 
$\sim$20\,K \citep[e.g.,][]{beckwith90,aw05,aw07b}, the integrated SMA flux 
density suggests a large total mass of gas and dust is present around IRAS 
04158+2805, $\sim$0.03\,M$_{\odot}$.  This value is in good agreement with the 
disk mass range inferred by \citet{glauser08}, $M_d \approx 
0.02$-0.04\,M$_{\odot}$.  Estimating masses for circumstellar material in this 
way is inherently uncertain (perhaps by an order of magnitude), due to the 
challenge of observationally constraining the optical properties of dust grains 
and the assumed mass conversion of a trace constituent (dust) to the presumably 
dominant species (molecular gas).  

Perhaps more compelling (and easier to interpret) is the unusual geometry that 
is observed for the dust structure around IRAS 04158+2805.  The enormous size 
of the scattered light nebulae noted by \citet[][2240\,AU diameter]{glauser08} 
and the submillimeter continuum emission presented here (1120\,AU diameter) is 
exceptionally rare for circumstellar disks 
\citep[e.g.,][]{pietu07,watson07,aw07,hughes08}, rivaled only by the 
circumbinary disk around UY Aur \citep{duvert98,close98,potter00}.  On the 
contrary, the observed size is fairly typical (if not on the small end) for 
larger-scale envelopes \citep[e.g.,][]{looney00,eisner05}.  

\begin{figure}
\epsscale{0.6}
\plotone{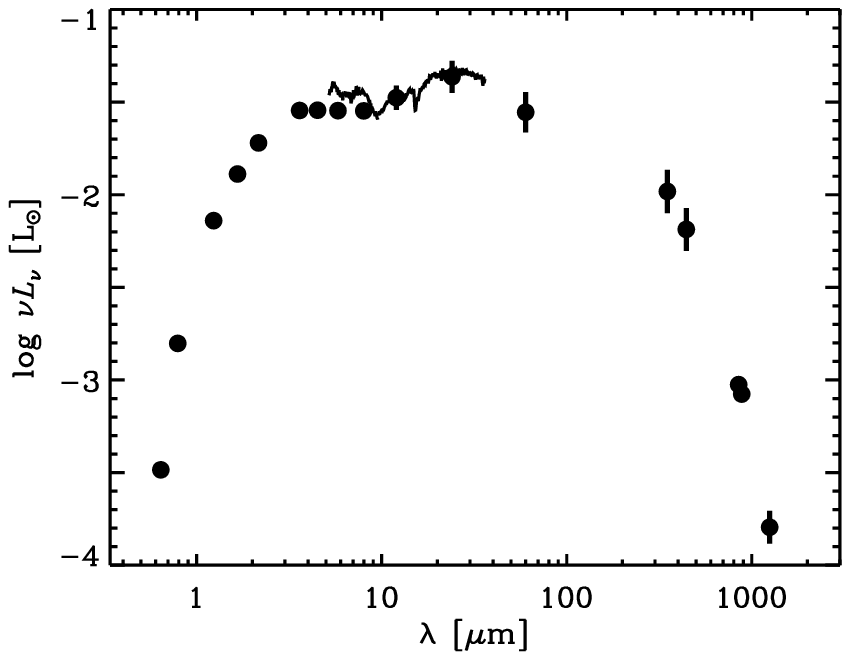}
\caption{The \iras spectral energy distribution.  Optical photometry was taken
from \citet{luhman98}, infrared data from the 2MASS Point Source Catalog
\citep{cutri03}, \citet{luhman06b}, and \citet{beichman86}, the mid-infrared 
{\it Spitzer} IRS spectrum from \citet{furlan08}, the submillimeter data from 
measurements presented here, and the 1.3\,mm flux density from 
\citet{motte01}.  See \citet{furlan08} for a more detailed discussion of the 
mid-infrared spectral features.  \label{sed}}
\end{figure}

Moreover, our resolved observations of the submillimeter continuum reveal that 
a large central region (perhaps $\sim$500\,AU in diameter) has significantly 
diminished intensity at 883\,$\mu$m, suggesting a ring-like or toroidal 
geometry for the dust structure.  Similar geometries have been inferred for a 
handful of circumstellar disks, and a variety of underlying causes are 
possible.  The so-called ``transition" disks have their inner regions (out to a 
few tens of AU in radius) largely cleared of observable material, presumably by 
disk evolution processes like particle growth, photoevaporation, or dynamical 
interactions with a young planetary system 
\citep[e.g.,][]{pietu06,hughes07,brown08}.  However, given the large spatial 
scale of the central depression and the absence of a corresponding dip in the 
infrared part of the \iras SED, as shown in Figure \ref{sed}, this scenario is 
unlikely.  

Alternatively, extremely dense disks with edge-on orientations can potentially 
render even the submillimeter emission optically thick, resulting in a central 
depression in its brightness distribution \citep[e.g.,][]{wolf08}.  Although 
the \iras dust structure is not oriented edge-on, perhaps longer pathlengths 
through the exceptionally large structure can provide compensating column 
densities.  If the bulk of the submillimeter emission is optically thick, the 
spectrum should have a spectral index $\alpha = 2$, where $F_{\nu} \propto 
\nu^{\alpha}$ \citep{beckwith91}.  Power-law fits to the submillimeter 
photometry are significantly steeper than the optically thick case, with 
$\alpha = 2.7\pm0.3$ (0.45-1.3\,mm; $\alpha = 3.8\pm0.6$ for 0.85-1.3\,mm).  At 
these wavelengths, such a steep spectrum is only noted for a few percent of 
Class II sources in Taurus and $\rho$ Oph, but is more common (25-35\%) in the 
envelopes around Class I objects \citep{aw05,aw07b}.  The observed steep 
submillimeter SED would be difficult to produce unless a large fraction of the 
dust was optically thin at those wavelengths.  As noted by \citet{wolf08}, a 
clear test of this possibility can be made with resolved continuum observations 
at $\lambda > 883$\,$\mu$m.  In such data, the material should become optically 
thick only at significantly smaller radii than at 883\,$\mu$m, resulting in a 
more centrally concentrated emission profile (or, equivalently, the null in the 
visibilities should be detected at a comparatively longer deprojected 
baseline).  Note that if the material inside this central depression is 
optically thick at 883\,$\mu$m, an enormous reservoir of mass is not accounted 
for in the above estimate.  

A third possibility for explaining the submillimeter continuum depression is 
that the emitting dust resides in a circumbinary ring.  The material around 
close binary stars is expected to be dynamically cleared on scales similar to 
the physical separation of the stellar components 
\citep[e.g.,][]{jensen96,guilloteau99}.\footnote{Recent observations suggest 
that the circumbinary scenario may be the more appropriate explanation in some 
cases for the diminished central emission in both transition \citep{ireland08} 
and dense, edge-on disks \citep{guilloteau08}.}  Numerical simulations of this 
dynamical interaction suggest that a circumbinary disk gap would be cleared out 
to a radius $\sim$1.5-3$\times$ the projected semimajor separation of the 
stellar components, and the individual circumstellar disks would be truncated 
at a radius $\sim$0.2-0.5$\times$ the projected semimajor separation 
\citep[e.g.,][]{artymowicz94}.  If the 883\,$\mu$m continuum depression noted 
here is caused by such clearing, we would expect the projected binary 
separation to be $\sim$90-180\,AU ($\sim$0\farcs6-1\farcs3) for a reasonable 
range of eccentricities, and the remnant individual circumstellar disks might 
be truncated at radii of $\sim$20-90\,AU.  This configuration is similar to the 
UY Aur system, where a ring with $\sim$20\arcsec\ diameter is detected in both 
scattered light and CO around a roughly equal-mass binary with a similar 
projected separation \citep{duvert98,close98,potter00}.  The primary difference 
is in the masses involved; the stellar mass in the \iras system is 
$\sim$4$\times$ less than for UY Aur, but the circumstellar mass (or rather 
850\,$\mu$m luminosity) is $\sim$4$\times$ higher.  While this possibility is 
certainly appealing in its reconciliation of both the information on the 
central source (see \S 4.1) and the structure of the gas and dust emission, 
testing the multiplicity of \iras will be a challenge given the obscuration of 
the central region by the dust structure.  Decomposing high resolution spectra 
of scattered starlight may represent the best opportunity in this case. 

The \iras SED, shown in Figure \ref{sed}, exhibits a strong infrared excess 
with a flat or slightly rising slope from 2-25\,$\mu$m.  A variety of 
solid-state absorption features are noted in higher resolution infrared 
spectra, including H$_2$O ice (3\,$\mu$m; not shown), silicates (10\,$\mu$m), 
and CO$_2$ ice \citep[15\,$\mu$m;][]{beck07,furlan08}.  The bright and steep 
submillimeter spectrum has already been discussed above.  While all of these 
SED properties are not necessarily inconsistent with a large, cold, highly 
inclined circumstellar disk 
\citep[e.g.,][]{menshchikov97,chiang99,pontoppidan05}, taken together they are 
more common for Class I/Flat-Spectrum objects that harbor a remnant accretion 
envelope \citep[e.g.,][]{whitney03,watson04,boogert04,pontoppidan07}.  
\citet{furlan08} demonstrate that the \iras SED features are strikingly similar 
to those for several other Class I sources in Taurus (see their Fig.~2) and can 
be successfully reproduced in detail with an envelope model (although the new 
observations of this source warrant a modification of their geometric 
assumptions).  So, all of the observational evidence presented here is also 
consistent with a flattened envelope structure around IRAS 04158+2805.  Such a 
flattened envelope geometry is predicted by molecular cloud core collapse 
models that incorporate rotation or magnetic fields \citep{terebey84,galli93}, 
and have been clearly detected in other cases, albeit on significantly larger 
spatial scales than noted here \citep[e.g.,][]{looney07}.

In this scenario, a partially evacuated outflow cavity at the center of the
flattened envelope may explain the central depression in the submillimeter
continuum data \citep[e.g.,][their Models 3 or 4]{whitney03}.  
\citet{glauser08} indicate that \iras drives a substantial H$\alpha$ jet
oriented in the north-south direction, perpendicular to the observed dust
structure.  The velocity-integrated CO $J$=3$-$2 morphology does resemble the
bases of some molecular outflows \citep[e.g.,][]{arce06}, but the absence of
both a velocity gradient in the proposed flow direction (north-south) and
larger scale molecular outflow signatures \citep{bontemps96,gomez97} are
difficult to reconcile with that interpretation.  Nevertheless, some
contribution to the CO emission from an outflow and even the envelope (i.e.,
motions unrelated to Keplerian rotation) would complicate the observed
spatio-kinematics.  The dynamical estimate of $M_{\ast}$ discussed above
should be treated with caution with regards to this possibility.  Of course, 
the envelope and binary ideas are not mutually exclusive, and dynamical 
clearing of the inner envelope and disks is still a viable alternative.  

The images in Figures \ref{images}c and \ref{chanmaps} demonstrate that, 
unlike the submillimeter continuum, the CO emission is centrally concentrated; 
most of the line emission is from radii inside the toroidal structure traced in 
the continuum.  While this may suggest that the bulk of the line emission is 
generated in the disk(s) interior to that structure, it is difficult to provide 
a reliable origin without a more complete physical model for the \iras 
environment.  Future studies of this source should focus on searching for 
definitive molecular outflow signatures, resolving dense gas tracers that are 
commonly associated with envelopes \citep[e.g.,][]{jorgensen07}, observing CO 
transitions at higher spectral resolution to provide a better tracer of the 
velocity field, and reconciling the optical/near-infrared scattered light 
observations with the submillimeter results in an effort to converge on a 
consistent model of the local gas and dust structure.

Considering the low stellar mass range implied for IRAS 04158+2805, it is 
difficult to reconcile the large mass reservoir and size of this environment 
with any formation scenario that relies on it being dynamically ejected from a 
more massive system \citep[e.g.,][]{reipurth01,bate03}.  Rather, the observed 
gas and dust structure is more consistent with the standard scenario, where an 
accretion disk and perhaps a remnant envelope developed around a single (or 
binary) source during the gravitational collapse of a molecular cloud core 
fragment.  Regardless of the precise value of $M_{\ast}$, \iras is a remarkable 
example of the fact that the standard picture for star formation worked out for 
higher-mass stars is also applicable at the low end of the stellar mass 
function.

\section{Summary}

We have presented new submillimeter observations of the molecular gas and dust 
around IRAS 04158+2805, a young low-mass object in the Taurus star-forming 
region.  The data reveal a complex and surprising environment, from which we 
infer the following:

1.  A high aspect ratio ($i \approx 62$\degr) dust structure with a projected 
diameter of $\sim$1120\,AU and estimated (gas+dust) mass of 
$\sim$0.03\,M$_{\odot}$ is noted in optical absorption and submillimeter 
continuum emission images.  The resolved 883\,$\mu$m data clearly demonstrate 
a central depression in the emission, suggesting a ring-like geometry with an 
inner radius of $\sim$250-300\,AU.  A variety of structures that may explain 
the observations are investigated, with a flattened envelope or circumbinary 
ring perhaps the most likely.

2.  Spatially resolved CO $J$=3$-$2 line emission shows a velocity gradient 
along the major axis of the dust structure.  Assuming the gas traced by this 
transition is in Keplerian orbital motion around a central source, the 
spatio-kinematic properties of the emission are used to estimate a central 
mass, $M_{\ast} \approx 0.3$\,M$_{\odot}$ (with $\sim$50\% uncertainty).  This 
estimate is roughly a factor of 2 higher than the mass range inferred from a 
reasonable effective temperature range and pre-main-sequence models, 
$\sim$0.09-0.16\,M$_{\odot}$, but the significant uncertainties on both 
measurements do not formally rule out their mutual consistency.  However, a 
roughly equal-mass binary system represents an intriguing possibility in this 
case, as it would naturally account for this possible $M_{\ast}$ discrepancy, a 
very high luminosity, and the ring-like geometry of the dust structure.

3.  The large spatial extent and mass of the complex gas and dust environment 
surrounding \iras would be difficult to retain in the dynamical ejection models 
used to explain the formation of such low-mass objects.  Instead, this case 
appears to illustrate that the standard gravitational collapse models applied 
to stars with masses up to $\sim$2-3\,M$_{\odot}$ can be extrapolated down to 
objects with $M_{\ast} \sim 0.1$-0.3\,M$_{\odot}$.  More detailed models of the 
\iras gas and dust structure and specialized observations are recommended.

\acknowledgments 
We thank F.~M{\'{e}}nard and G.~Duch{\^{e}}ne for encouraging 
our interest in this source, and Elise Furlan for kindly providing the 
mid-infrared spectrum.  We are grateful to an anonymous referee for a helpful,  
constructive review and for emphasizing the potential circumbinary nature of 
IRAS 04158+2805.  The SMA is a joint project between the Smithsonian 
Astrophysical Observatory and the Academia Sinica Institute of Astronomy and 
Astrophysics and is funded by the Smithsonian Institution and the Academia 
Sinica.  Support for this work was provided by NASA through Hubble Fellowship 
grant \#HF-01203.01-A awarded by the Space Telescope Science Institute, which 
is operated by the Association of Universities for Research in Astronomy, Inc., 
for NASA, under contract NAS 5-26555.  MCL and KNA acknowledge support for this 
work from NSF grant AST-0407441.

\clearpage

\end{document}